\documentclass[12pt]{article}
\usepackage{amsmath}
\usepackage{amssymb}
\usepackage{graphicx}
\usepackage{enumerate}
\usepackage{natbib}
\usepackage{url} % not crucial - just used below for the URL 

\newcommand{\blind}{1}

\usepackage{hyperref}

\begin{document}

\def\spacingset#1{\renewcommand{\baselinestretch}%
{#1}\small\normalsize} \spacingset{1}

%%%%%%%%%%%%%%%%%%%%%%%%%%%%%%%%%%%%%%%%%%%%%%%%%%%%%%%%%%%%%%%%%%%%%%%%%%%%%%

\if1\blind
{
  \title{\bf Combining Heterogeneous Spatial Datasets with Process-based Spatial Fusion Models: A Unifying Framework}
  \author{Craig Wang\hspace{.2cm}\\
    Department of Mathematics, University of Zurich\\
    and \\
    Reinhard Furrer \\
    Department of Mathematics, University of Zurich \\
    Department of Computational Science, Unversity of Zurich\\
    and \\
    for the SNC Study Group}
  \maketitle
} \fi

\if0\blind
{
  \bigskip
  \bigskip
  \bigskip
  \begin{center}
    {\LARGE\bf Combining Heterogeneous Spatial Datasets with Process-based Spatial Fusion Models: A Unifying Framework}
\end{center}
  \medskip
} \fi

\bigskip
\begin{abstract}
In modern spatial statistics, the structure of data that is collected has become more heterogeneous. Depending on the type of spatial data, different modeling strategies for spatial data are used. For example, a kriging approach for geostatistical data; a Gaussian Markov random field model for lattice data; or a log Gaussian Cox process for point-pattern data. Despite these different modeling choices, the nature of underlying scientific data-generating (latent) processes is often the same, which can be represented by some continuous spatial surfaces. In this paper, we introduce a unifying framework for process-based multivariate spatial fusion models. The framework can jointly analyze all three aforementioned types of spatial data (or any combinations thereof). Moreover, the framework accommodates different conditional distributions for geostatistical and lattice data. We show that some established approaches, such as linear models of coregionalization, can be viewed as special cases of our proposed framework. We offer flexible and scalable implementations in R using Stan and INLA. Simulation studies confirm that the predictive performance of latent processes improves as we move from univariate spatial models to multivariate spatial fusion models. The introduced framework is illustrated using a cross-sectional study linked with a national cohort dataset in Switzerland, we examine differences in underlying spatial risk patterns between respiratory disease and lung cancer.
\end{abstract}

\noindent%
{\it Keywords:}  Data fusion; Bayesian modeling; Gaussian process; Large dataset; Change of support problem; Stan; INLA.

\spacingset{1.5} % DON'T change the spacing!
\section{Introduction}
\label{sec:intro}
In statistics, spatial models are useful when residuals exhibit correlation in space after accounting for known covariates in a regression-type setting. Spatial data has long been classified into three categories, namely geostatistical (point-level) data, lattice (area-level) data, and point-pattern data \citep{cressie2015}. Due to reasons such as measurement method constraints and privacy considerations, different types of spatial data may be collected in different settings. Depending on the data type, different statistical models that capture the residual spatial correlation are used. In a nutshell, (1) geostatistical data are commonly collected in environmental science. For example, rainfall at weather stations \citep{Kyriakidis2001}, where the exact geo-coordinates for each observation are known. The strength of dependency is a function of the distance separation between two locations. Kriging can be used to model a smooth surface. (2) Lattice data can be either gridded or irregularly aligned, and occur in the form of aggregated observation over areas. They are often collected in epidemiology, such as disease prevalence of each district \citep{Chammartin2016}. Another source of lattice data is from measuring instruments, such as satellites where the spatial resolution is intrinsically limited, resulting in gridded observations. Gaussian Markov random field (GMRF) models such as conditionally autoregressive models are typically used to capture the spatial dependency between neighboring areas. Finally, (3) there is point-pattern data, where the locations themselves are stochastic. They are used to model epidemiological data of case locations \citep{Gatrell1996} or other event locations such as epicenter of earthquakes \citep{Ogata1998}. One approach to model such data is using a Poisson process, where the intensity function may depend on observed covariates. Despite having different modeling strategies for different types of data, a common purpose of all the aforementioned statistical models is to capture the residual spatial dependency between different observations. A natural way to do this is using Gaussian processes to model continuous spatial surfaces, which represent the underlying scientific process that drives the response variables together with observed covariates.

There are already attempts to analyze lattice data and point-pattern data using Gaussian process-based models. For example, \cite{Kelsall2002} modeled aggregated disease counts using a Gaussian process approach. The authors derived analytical approximations to area-level Poisson mean and produced continuous underlying relative risk functions using Markov chain Monte Carlo (MCMC). Point-pattern data can be linked to Gaussian process with a log-Gaussian Cox process (LGCP) \citep{Moller1998}. Instead of having a fixed intensity at each location, the intensity is modeled as a log-Gaussian process, yielding a doubly stochastic Poisson process. In modern spatial statistics, researchers are dealing with increased heterogeneity in the structure of spatial data that is collected. Different data sources may contain overlapping information concerning the same research questions. Recently, there are several approaches in the literature that proposed to use Gaussian process as a basis to fuse spatial data of different types. \cite{Moraga2017} proposed a model to analyze spatial data available in both geostatistical and lattice type, with the same set of covariates and a response variable observed at two different spatial resolutions. Their computation is made efficient using integrated nested Laplace approximations (INLA) \citep{Rue2009}. \cite{Wilson2018} extended the work by allowing non-Gaussian response variables. This opens up possibilities of modeling count data which commonly occurs in epidemiological settings. \cite{Shi2017} proposed a fixed rank kriging-based fusion model to combine multiple lattice-type remote sensing datasets. Other works on spatial fusion models  \citep{Berrocal2010,McMillan2010, Sahu2010} also implemented efficient algorithms for specific model structure introduced in their applications. In general, spatial fusion models are challenged by their flexibility and computational efficiency. There exists a trade-off between them, i.e. more flexible modeling structure comes with a higher computational cost. For example, as shown in \cite{Wilson2018}, their fusion model with normally-distributed response variables took an order of minutes using INLA to compute. A more flexible modeling structure for Poisson-distributed response took several weeks using Hamiltonian Monte Carlo-based inference method. 

In this paper, we extend previous works in those two aspects. In terms of flexibility, our framework incorporates an additional data type, namely point-pattern data. To the best of our knowledge, this is the first spatial fusion framework that incorporates all three types of spatial data. We additionally allow arbitrary combinations of those three data types in multivariate settings. We propose a unifying framework that includes the features of several well-established models, such as linear models of co-regionalization (LMC) \citep{Wackernagel2003,MacNab2016}, spatial factor model \citep{Wang2003}, and shared component model \citep{Held2001}. In terms of computational cost, we implement a fully Bayesian-based approach using Stan modeling language \citep{stan2017}. As a more efficient alternative, we offer INLA-based implementation which significantly reduces computation time from hours to minutes for thousands of observations. Last but not least, we benchmark the performance of these two implementations in terms of prediction and parameter estimation in simulation studies.

The rest of this paper is structured as follows: Section~\ref{sec:mod} introduces the unifying framework and explicitly show its link to existing spatial models. Section~\ref{sec:inf} discusses implementation strategies in Stan and INLA. Section~\ref{sec:sim} illustrates the framework using two simulated scenarios and an analysis on epidemiological datasets, as well as comparing the performance of our two implementations. Finally, we end with a summary followed by some discussion on identifiability problems and research outlook in Section~\ref{sec:dis}.

\section{Process-based Spatial Fusion Model}
\label{sec:mod}
\subsection{The Unifying Framework}
\begin{figure}
\includegraphics[width = \textwidth]{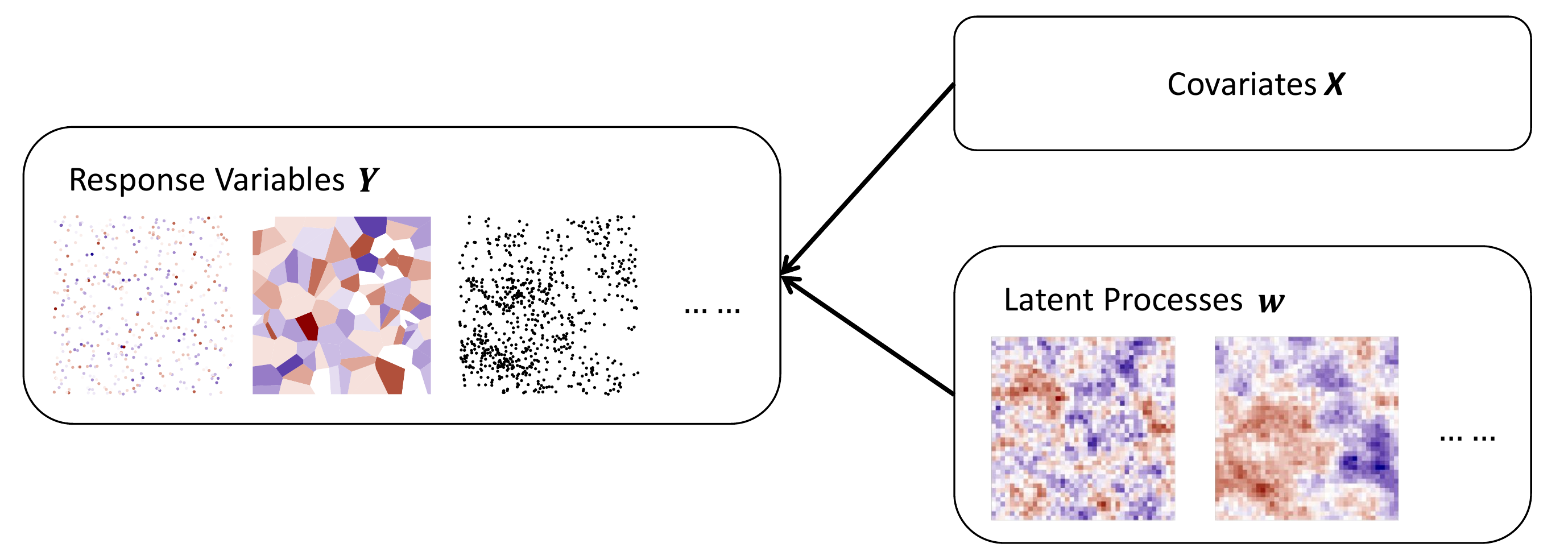}
\caption{A graphical formulation of the spatial fusion model framework, consisting of multiple response variables of different type and multiple latent processes.}
\label{fig:abs}
\end{figure}

For $j=1,\dots, \ell$, we let $\boldsymbol{Y}_j(\cdot)$ denote the $j$th response variable with $n_j$ observations, with a conditional distribution that belongs to the exponential family. Each of the $\ell$ response can take any of the following data types: i) geostatistical data, observed at locations $\boldsymbol{s}_j \in D \subseteq \mathcal{R}^2$; ii) lattice data observed at areas $\boldsymbol{a}_j \subset D$; or iii) point-pattern data that has been discretized to regular fine grid containing mostly zeros or ones, observed at gridded locations $\boldsymbol{v}_j \in D$, where $\boldsymbol{Y}_j(\boldsymbol{v}_j)$ denotes the number of events in the grid cell containing $\boldsymbol{v}_j$. Further, we let $\boldsymbol{X}_j(\cdot)$ denote a full (column) rank $n_j \times p$ matrix of spatially-referenced covariates that are observed at the same spatial units as the corresponding response variables, $\boldsymbol{\beta}_j$ denote a vector $p \times 1$ of fixed effect coefficients. We assume there is a $q\times 1$ vector of zero-mean, unit variance, independent latent Gaussian processes $\boldsymbol{w}(\cdot)$ having a $\ell \times q$ design matrix $\boldsymbol{Z}$, i.e. $\boldsymbol{Z}_j\boldsymbol{w}(\cdot)$ is the $j$th linear combination of Gaussian processes. Each Gaussian process is parameterized by its own covariance function. Finally, non-linear operator $B_j(\cdot)$ subsets and aggregates some components of $\boldsymbol{Z}_j\boldsymbol{w}(\cdot)$ such that it matches the spatial resolution of the corresponding response variable. Overall, the framework can be formulated as
\begin{equation}
g_j(\mathbb{E}[\boldsymbol{Y}_j(\cdot) \vert \boldsymbol{\beta}_j, \boldsymbol{Z}_j, \boldsymbol{w}(\cdot)]) = \boldsymbol{X}_j(\cdot)\boldsymbol{\beta}_j + B_j(\boldsymbol{Z}_j\boldsymbol{w}(\cdot)),
\end{equation}
where $g_j(\cdot)$ is a link function that corresponds to the conditional distribution of $\boldsymbol{Y}_j(\cdot)$. Fig.~\ref{fig:abs} outlines a graphical formulation of the framework.

Change of support problems \citep{Gotway2002} arise when lattice data needs to be modeled. We only observe aggregated information while the underlying process is continuous. We employ a sampling-points approximation approach to stochastic integrals \citep{Gelfand2001,Fuentes2005} for aggregating latent processes. Let $\boldsymbol{s}'$ denote the set of all sampling points and each area contains $H$ sampling points. For the $i$th area $\boldsymbol{a}_{ji}$ in $j$th response, under a linear link function we obtain
\begin{equation}
\label{eq:linear}
w(\boldsymbol{a}_{ji}) = \vert \boldsymbol{a}_{ji} \vert^{-1} \int_{\boldsymbol{u}\in \boldsymbol{a}_{ji}}w(\boldsymbol{u})d\boldsymbol{u} \approx \frac{1}{H}\sum_{\boldsymbol{s}'\in \boldsymbol{a}_{ji}}w(\boldsymbol{s}').
\end{equation}
When a nonlinear link function is used for a response variable, the aggregation in \eqref{eq:linear} will result ecological bias \citep{Greenland1992}. For a general link function $g_j(\cdot)$, we have the following approximation,
\begin{equation}
\label{eq:nonlinear}
w(\boldsymbol{a}_{ji}) = g_j\left(\vert \boldsymbol{a}_{ji} \vert^{-1} \int_{\boldsymbol{u}\in \boldsymbol{a}_{ji}} g_j^{-1}\left(w(\boldsymbol{u})\right)d\boldsymbol{u}\right) \approx g_j\left(\frac{1}{H}\sum_{\boldsymbol{s}'\in \boldsymbol{a}_{ji}}g_j^{-1}\left(w(\boldsymbol{s}')\right)\right).
\end{equation}
Typically, a small $H$ is chosen to balance the trade-off between computational efficiency and model accuracy \citep{Fuentes2005,Liu2011}.

Albeit the latent processes have a continuous index, we work with a finite set of locations in practice.
The set of locations $\mathcal{U}$ to be modeled in the latent processes $\boldsymbol{w}(\cdot)$ comprise the locations where geostatistical data are observed, the locations of sampling points for lattice data and gridded locations for point-pattern data. The non-linear operator $B_j(\cdot)$ takes a different form for different data types. For geostatistical and point-pattern data, the non-linear operator $B_j(\cdot)$ subsets $\boldsymbol{Z}_j\boldsymbol{w}(\cdot)$ to the corresponding locations of the $j$th response variable. For lattice data and a linear link function, $B_j(\cdot)$ subsets $\boldsymbol{Z}_j\boldsymbol{w}(\cdot)$ to the sampling point locations and aggregates them to the corresponding areas by taking averages. With non-linear link functions, $B_j(\cdot)$ first applies an inverse link function and then aggregates.

\subsection{Link to Existing Models}
Our proposed unifying framework utilizes elements from existing literature and combines them to create a flexible yet efficient spatial fusion model framework. As a result, there are some strong links in terms of model structure between this framework and some established methods in spatial statistics. At the same time, they share the same potential identifiability issues. 

In univariate settings, the unifying framework allows us to model each type of spatial data individually with a latent Gaussian process. When we have geostatistical data, it results in a geostatistical regression \citep{cressie2015}. With Poisson-distributed lattice data, we obtain a sampling-points approximation to the model used in \cite{Kelsall2002}, which is an alternative modeling strategy to Besag-York-Mollé model \citep{BYM}. With point-pattern data, we obtain a discretized LGCP \citep{Moller1998}.

In multivariate geostatistical data settings, the design matrix $\boldsymbol{Z}$ plays a pivotal role in the identifiability of model parameters. When the number of independent Gaussian processes is less than the number of responses $q < \ell$, we obtain a spatial factor model \citep{Wang2003}. The latent spatial factors are assumed to have zero-mean unit-variance Gaussian processes, such that $\boldsymbol{Z}$ controls the variance (partial sill) of latent processes. When $q = \ell$, we obtain a general LMC framework \citep{Wackernagel2003,Schmidt2003}. A similar LMC framework also exists for lattice data \citep{MacNab2016}. Identifiability issues occur in the LMC since the number of latent values to be estimated in the latent processes is equal to the total number of observations in response variables. Additional spatial hyper-parameters, and fixed-effect coefficients also need to be estimated. For this reason, regularization is done via one of the following: 1) empirical Bayes method by fixing some of the hyper-parameters; 2) choosing informative prior distributions in Bayesian models; or 3) using a lower triangular matrix for $\boldsymbol{Z}$ \citep{Schmidt2003}. In cases of $q > \ell$, we acquire a similar model structure as shared component models \citep{Held2001} for Gaussian processes, where multiple outcomes have their own latent spatial components plus some shared spatial components. In this setting, the values in $\boldsymbol{Z}$ need to be even further constrained to avoid identifiability issues \citep{Held2001}.

Our framework is also linked to other process-based spatial data fusion models, which combine geostatistical and lattice data types. When we let the response variables represent the same information but have different data types, we obtain the model presented in \cite{Wilson2018}, where an explicit relationship is used to link multiple response variables. If we further allow different information to be represented in the response variables, we reach the generalized spatial fusion model framework proposed in \cite{Wang2018}. 

To the best of our knowledge, there is no existing approach or implementation that jointly models all three types of spatial data in a multivariate framework. With those links to the existing approaches, our framework extends upon them by combining different features and enhance the overall flexibility of spatial fusion models.

\section{Model Implementations}
\label{sec:inf}
It is well known that fitting full Gaussian processes in Bayesian models is computationally expensive in both univariate and multivariate settings. Marginalized and conjugate Gaussian process models dramatically save computation time but they are only available when fitting geostatistical data with normally-distributed outcomes \citep{Banerjee2014,zhang2019}. There exist several approaches to reduce the computational burden, such as low rank \citep{Cressie2007, Banerjee2008, Stein2008} and sparse \citep{Furrer2006, Rue2009, Datta2016} methods. Some of those approaches are utilized in existing spatial fusion models. \cite{Shi2017} adapted the spatial basis function approach from fixed rank kriging \citep{Cressie2007}. \cite{Moraga2017} used integrated nested Laplace approximations \citep{Rue2009}. \cite{Wang2018} exploited the nearest neighbor Gaussian process (NNGP) \citep{Datta2016}. In this paper, we offer two efficient implementation strategies for the unifying spatial fusion model framework. The first strategy follows an adaptation of NNGP implementation in \cite{Wang2019}. The second strategy follows \cite{Wilson2018} to use INLA, with additional approximations for non-linear link functions.

\subsection{Implementation using NNGP}
Fitting full Gaussian processes in a Bayesian hierarchical model is costly, therefore we seek for efficient methods to speed up the inference. An NNGP implementation approximates a full Gaussian process by assuming that latent variables in the Gaussian process are conditionally independent given their neighborhood sets, hence introducing sparsity in the precision matrix. \cite{Datta2016} showed that it significantly reduces computation time in geostatistical models while yielding results close to a full Gaussian process-based inference. In our implementation, we let each latent spatial process $w(\boldsymbol{\cdot})$ following an independent NNGP. Let $w_\mathcal{U}$ denote a latent process on the set of locations $\mathcal{U}$, then the NNGP likelihood according to \cite{Datta2016} can be written as 
\begin{equation}
\label{eq:nngp}
p(w_\mathcal{U}) = \prod_{i=1}^{n_\mathcal{U}} \text{N}\Big(w(\boldsymbol{u}_i)\mid C_{\boldsymbol{u}_i, \mathcal{N}(\boldsymbol{u}_i)}C^{-1}_{\mathcal{N}(\boldsymbol{u}_i)}w_{\mathcal{N}(\boldsymbol{u}_i)}, C_{\boldsymbol{u}_i,\boldsymbol{u}_i}-C_{\boldsymbol{u}_i, \mathcal{N}(\boldsymbol{u}_i)}C^{-1}_{\mathcal{N}(\boldsymbol{u}_i)}C_{\boldsymbol{u}_i, \mathcal{N}(\boldsymbol{u}_i)}\Big),
\end{equation}
where $\mathcal{N}(\boldsymbol{u}_i)$ is the set of $\max(i-1,m)$ nearest neighbors from $\{\boldsymbol{u}_1, \boldsymbol{u}_2,\dots, \boldsymbol{u}_{i-1}\}$ for location $\boldsymbol{u}_i$ with a fixed constant $m$, $C_{\boldsymbol{u}_i, \mathcal{N}(\boldsymbol{u}_i)}$ is the cross-covariance matrix between the latent process $w(\boldsymbol{u}_i)$ and its neighbors $\mathcal{N}(\boldsymbol{u}_i)$, $C_{\mathcal{N}(\boldsymbol{u}_i)}$ is the covariance matrix of $w_{\mathcal{N}(\boldsymbol{u}_i)}$, and $C_{\boldsymbol{u}_i,\boldsymbol{u}_i}$ is the variance of $w(\boldsymbol{u}_i)$. The variance and covariance matrices are parameterized by spatial hyperparameters. 

The full Bayesian hierarchical model is then implemented using Stan modeling language via the \textbf{rstan} package \citep{rstan2016} in R \citep{R2017}, consisting of likelihoods for each outcome variable and NNGP, as well as priors for fixed effect coefficients and spatial hyperparameters. Stan implements the No-U-Turn sampler \citep{Homan2014} based on Hamiltonian Monte Carlo, which provides efficient means of conducting full Bayesian inference for complex hierarchical structures. 

\subsection{Implementation using INLA}
Although the computational efficiency can be improved by using NNGPs instead of full Gaussian processes, it is still not feasible to fit multiple latent processes with more than thousands of locations in $\mathcal{U}$. Therefore,  we implement an alternative strategy using INLA.

Over a fixed set of locations, a Gaussian process is equivalent to a Gaussian random field (GRF). \cite{Lindgren2011} established a connection between GRFs and GMRFs through a stochastic partial differential equation approach, where a GRF can be approximated by triangulating the spatial domain and using a weighted sum of basis functions as
\begin{equation}
\label{eq:inla}
w_\mathcal{U} \approx \sum_{k=1}^{m} r_k\phi_k,
\end{equation}
where $m$ is the number of points in the triangulation, $r_k$ are Gaussian distributed weights and ${\phi_k}$ are basis functions. The weights $\boldsymbol{r} = [r_1,r_2,\dots,r_t]$ forms a GMRF with sparse precision matrix which makes computation efficient. In this approach, the covariance function must be a member of the Mat\'{e}rn family defined as
\begin{equation}
C_{\boldsymbol{u}_i, \boldsymbol{u}_j} = \frac{\sigma^2}{2^{\nu-1}\Gamma(\nu)}(\sqrt{2\nu}||\boldsymbol{u}_i - \boldsymbol{u}_j||/\phi)^\nu K_\nu (\sqrt{2\nu}||\boldsymbol{u}_i - \boldsymbol{u}_j||/\phi) ,
\end{equation}
where $||\boldsymbol{u}_i - \boldsymbol{u}_j||$ is the Euclidean distance between $\boldsymbol{u}_i$ and $\boldsymbol{u}_j$, $K_v$ is the modified Bessel function of second kind with order $\nu$, $\sigma^2$ is the partial sill and $\phi$ relates to the spatial range, with $\nu \in (0,1]$ being the smoothness parameter. The approximation in Eq.~\eqref{eq:inla} can be written as $w_\mathcal{U} \approx A\boldsymbol{r}$, where $A$ is a projection matrix that maps a GMRF defined on the triangulation mesh nodes to the observations' locations. 

The key to implementing the spatial fusion models in INLA lies within the projection matrix, with a different structure required for each data type \citep{inlabook}. For geostatistical data, the $i$th row of the projection matrix corresponds to the $i$th location, it is filled with zeros except where 1) the location is on the $j$th vertex, then the $j$th column is 1 or 2) the location is within a triangulation area, then three cells get values based on a mixture of barycentric based weights from three neighboring vertices of the triangulation. For lattice data, we construct a projection matrix that links the $i$th area with the mean value of the GRF at mesh nodes which falls into the $i$th area. If the link function is linear, increasing the mesh density will increase the number of mesh nodes that falls into each area hence better approximate the average. However, for non-linear link functions, it is preferable to have less-dense mesh due to Jensen's inequality \citep{jensen1906}, which states
\begin{equation}
\frac{1}{H}\sum_{\boldsymbol{s}'\in \boldsymbol{a}_{ji}}g_j^{-1}\left(w(\boldsymbol{s}')\right) \gtrapprox g_j^{-1}\left(\frac{1}{H}\sum_{\boldsymbol{s}'\in \boldsymbol{a}_{ji}}w(\boldsymbol{s}')\right)
\end{equation}
for the $i$th area in the $j$th response. The approximation is better when there is only a smaller number of mesh nodes within each area \citep{follestad2003}. Finally, for the point-pattern data, we use an augmentation approach by \cite{Simpson2016}, which avoids discretizing the spatial domain into grid cells. The projection matrix is built as an identity matrix with dimension equal to the total number of mesh nodes, row-binded with a projection matrix that is constructed on observed locations in the same way as the geostatistical case. 

The final model fitting is done by stacking the projection matrices together and assigning appropriate priors using the \textbf{INLA} \citep{rINLA} package in R. Advances in \textbf{INLA} \citep{Martins2013} such as allowing multiple likelihoods and `copy' feature made this implementation possible.

\section{Illustrations}
\label{sec:sim}
In this section, we conduct two simulation studies and an analysis of epidemiological datasets to illustrate our proposed framework. All results are obtained in R version 3.5.0 \citep{R2017}, on a Linux server with 256GB of RAM and two Intel Xeon 6-core 2.5GHz processors. All R codes used in the simulation studies are provided in the supplementary material.

\subsection{Simulation Study One}
\label{subsec:sim1}
We are interested in modeling a single latent spatial process within a $[0,10]\times [0,10]$ square, using three spatial response variables with one being from each type. First, we simulate a zero-mean GRF on densely uniformly distributed locations with a covariance matrix $C\left(\cdot,\cdot;\sigma^2, \phi\right)$. We then sub-sample 200 locations to obtain the latent process at observed locations. For lattice observations, we divide the square into 100 Voronoi cells and compute aggregated GRF from all locations while accounting for ecological bias using Eq.~\eqref{eq:nonlinear}. In addition, we generate a covariate for geostatistical and lattice response by sampling from a standard normal distribution. Afterwards, we generate a normally-distributed geostatistical response at the same sampled locations and a Poisson-distributed lattice response for each area. For point-pattern observations, we simulate from the same GRF on a coarse 20$\times$20 grid, then exponentiate the values to obtain intensity at the grid cells, afterwards we generate Poisson point process using each intensity value multiplied by cell area and an offset term as the final intensity. In summary, the response variables are generated according to
\begin{align}
\boldsymbol{Y}_1 \mid \boldsymbol{\beta}_1,\boldsymbol{w},\tau_1^2 &\sim \text{N}\left(\boldsymbol{X}_1 \boldsymbol{\beta}_1 + B_1(\boldsymbol{w}), \tau_1^2\boldsymbol{I}\right), \nonumber\\
\boldsymbol{Y}_2 \mid \boldsymbol{\beta}_2,\boldsymbol{w} &\sim \text{Pois}\left(\exp(\boldsymbol{X}_2 \boldsymbol{\beta}_2 + B_2(\boldsymbol{w}))\right), \nonumber\\
\boldsymbol{Y}_3 \mid \boldsymbol{w} &\sim \text{Pois}\left(A\exp(B_3(\boldsymbol{w}))\right).
\end{align}
In the simulation, we use an exponential covariance function ($\nu = 0.5$), i.e. $C(\boldsymbol{u}_i,\boldsymbol{u}_j;\sigma^2,\phi) = \sigma^2 \mathrm{exp}(-||\boldsymbol{u}_i-\boldsymbol{u}_j||/\phi)$. The influence of varying sample sizes and spatial hyperparameters on predictive performance was investigated by \cite{Wang2018}, therefore we only consider a single setup by setting $\sigma^2 = 0.5$ and $\phi = 1$. In addition, we set $\boldsymbol{\beta}_1 = (1, 5)^\top, \boldsymbol{\beta}_2 = (1, 1.5)^\top$ and $\tau^2 = 1$. $A$ is the product of grid cell area and an offset term which takes value 0.25. 

We consider seven different model specifications within our proposed framework. (i - iii) three univariate models using a single data type each, namely one of geostatistical, lattice and point-pattern data, (iv - vi) three fusion models using different combinations of two data types, and (vii) a multivariate fusion model combining all three response variables. In Stan implementation, the intercepts and coefficients are assigned with independent $\textrm{N}(0,5^2)$ priors. The variance parameters $\sigma^2$ and $\tau^2$ are assigned with inverse Gamma prior $\textrm{IG}(2,1)$, which has a mean of one and undefined variance. For the spatial decay $\phi$, a zero-truncated normal prior $\textrm{N}(1, 3^2)$ is assigned. We use $m=5$ nearest neighbors and $H=5$ sampling points randomly selected within each area. We run 4 chains of 2,000 iterations with 1,000 warm-up samples, without thinning for each model. Multiple chain convergence is checked with potential scale reduction factors \citep{Brooks1998}. For INLA implementation, we use penalized complexity (PC) prior for Mat\'ern GRF \citep{Fuglstad2018} with $\alpha$ fixed at 1.5, corresponding to the exponential covariance function. In addition, we choose the median practical spatial range to be 2 (corresponds to the median of $\phi$ being 1) and the probability of $\sigma$ greater than 1.7 is 5\%, such that the allocated probability mass is closely matched with the priors in Stan. The rest of the priors in INLA are default options. The same data was modeled using both Stan and INLA implementations for comparison. Additionally, the simulation is repeated 100 times for INLA. We leave out the Stan implementation in the repetition part due to its long computation time.

\begin{figure}
\includegraphics[width = \textwidth]{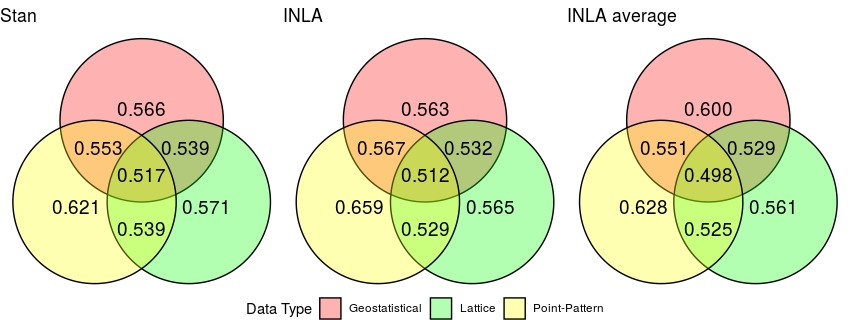}
\caption{Venn diagram of root mean squared prediction error for the unifying fusion framework fitted to each data type (and combinations thereof), using models implemented in Stan and INLA. Values in overlapping areas indicate results from models with multiple data types.}
\label{fig:sim1}
\end{figure}

We chose an additional 1600 sites to evaluate the predictive performance of models in terms of root mean squared prediction errors (RMSPE) under each scenario. The prediction sites are located at the centers of a $40\times 40$ grid that uniformly covers the sampling domain. Their predictive performance is shown in Figure~\ref{fig:sim1}. The first two venn diagrams show the RMSPEs for the simulated scenario under different models with Stan and INLA implementation. The last Venn diagram shows the average RMSPEs over 100 simulations for INLA. For both Stan and INLA implementations, the RMSPEs are smaller in multivariate fusion models compared to univariate process-based models. The joint modeling of all three types of spatial data has the lowest RMSPE on the prediction of the latent process at unobserved locations. Stan and INLA implementations produced comparable results, with the RMSPEs of Stan fall inside the ranges of repeated INLA simulations. 

\subsection{Simulation Study Two}
In the second simulation study, we focus on comparing the parameter estimates of a multivariate fusion model with three response variables of different types and two latent processes. We firstly simulate two independent zero-mean unit-variance GRFs uniformly distributed on the spatial domain of $[0,100]\times [0,100]$ square, then compute the sub-sampled and aggregated GRF for each response variable in the same way as in simulation one. Each response depends on the latent processes via the design matrix
\begin{equation}
\boldsymbol{Z} = \begin{bmatrix} 1.2 & 0 \\ 0.5 & 1.2 \\ 0 & 1 \end{bmatrix}.
\end{equation}
The first geostatistical response variable only depends on the first latent process, the second lattice response variable depends on both latent processes, while the third point pattern response variable depends only on the second latent process. The response variables are generated as follows,
\begin{align}
\boldsymbol{Y}_1 \mid \boldsymbol{\beta}_1,\boldsymbol{w},\tau_1^2 &\sim \text{N}\left(\boldsymbol{X}_1 \boldsymbol{\beta}_1 + B_1(\boldsymbol{Z}\boldsymbol{w}), \tau_1^2\boldsymbol{I}\right), \nonumber\\
\boldsymbol{Y}_2 \mid \boldsymbol{\beta}_2,\boldsymbol{w} &\sim \text{Pois}\left(\exp(\boldsymbol{X}_2 \boldsymbol{\beta}_2 + B_2(\boldsymbol{Z}\boldsymbol{w}))\right), \nonumber\\
\boldsymbol{Y}_3 \mid \boldsymbol{w} &\sim \text{Pois}\left(A\exp(B_3(\boldsymbol{Z}\boldsymbol{w}))\right).
\end{align}
where $\boldsymbol{Y}_1$ consists of 500 geostatistical observations, $\boldsymbol{Y}_2$ has 100 lattice observations and $\boldsymbol{Y}_3$ represents the number of events observed at each of 400 cells on a $20\times 20$ grid. In addition, we set $\boldsymbol{\beta}_1 = (3,5)^\top, \boldsymbol{\beta}_2 = (0.5,2)^\top, \phi_1 = 5$, $\phi_2 = 25$ and $\tau_1^2 = 0.5$. Since we have two latent processes in the simulation, using any of the univariate model or fusion model with two response variables can lead to identifiability problem. Hence, we estimate the parameters only using the unifying spatial fusion model with three responses only. The model and their prior specifications for both Stan and INLA are the same as in simulation one, except for the spatial range parameter. The prior for both $\phi_1$ and $\phi_2$ is zero-truncated $\textrm{N}(10,10^2)$ in Stan and PC prior with median practical spatial range $20$ in INLA (corresponds to the median of $\phi$ being 10).

\begin{figure}
\includegraphics[width = \textwidth]{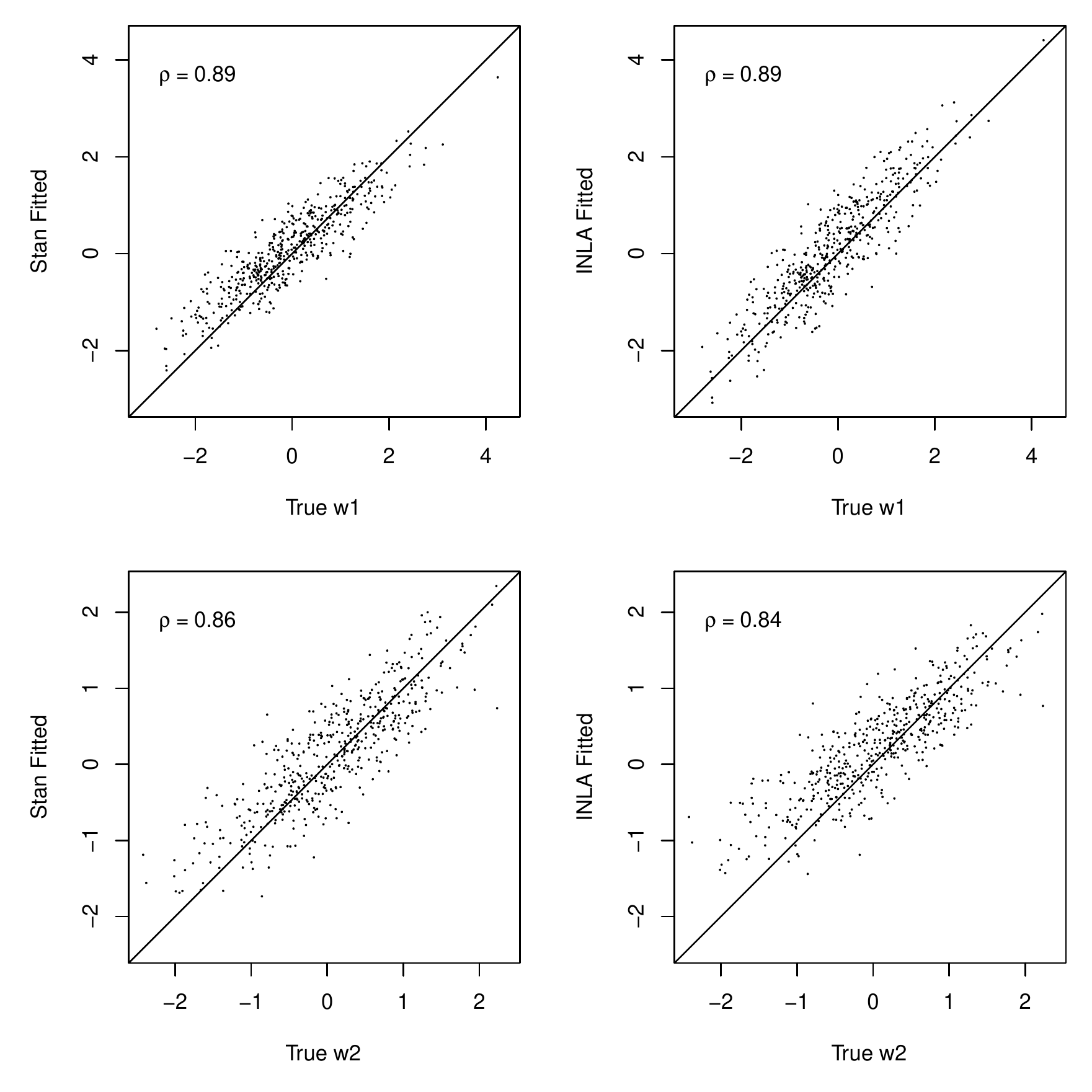}
\caption{True versus fitted latent process at locations with geostatistical observation. Pearson's correlation coefficients $\rho$ are displayed.}
\label{fig:sim2}
\end{figure}

The parameter estimates based on posterior medians and their 95\% posterior credible intervals for both implementations are summarized in Table~\ref{tab:par}. We obtained similar parameter estimates in both models. The PC prior in INLA penalizes complex structure in GRF hence tends to have a slightly over-estimated range. The posterior median of fitted latent processes at locations with geostatistical observations are shown in Fig~\ref{fig:sim2}. The root mean squared errors using Stan are 0.50 and 0.44 for the first and second latent process, compared to 0.54 and 0.48 using INLA. The computation time for the Stan implementation of the fusion model is 1.9 hours while it takes 11 minutes for INLA. 

\begin{table}
\caption{Parameter estimates and their 95\% posterior credible intervals (95\% CI) from the unifying spatial fusion model in simulation two with both Stan and INLA implementation.}
\label{tab:par}
\begin{tabular}{@{\extracolsep{30pt}}cccccc@{}}
\hline \hline
          & & \multicolumn{2}{c}{Stan}      & \multicolumn{2}{c}{INLA} \\ \cline{3-4} \cline{5-6}
               & True & Median  & 95\% CI          & Median  & 95\% CI \\ \hline
$\beta_{10}$ & 3.00 & 2.78 & (2.43, 3.11) & 2.78 & (2.41, 3.13) \\ 
  $\beta_{11}$ & 5.00 & 5.03 & (4.92, 5.13) & 5.01 & (4.91, 5.12) \\ 
  $\beta_{20}$ & 0.50 & 0.56 & (0.24, 0.84) & 0.61 & (0.3, 0.89) \\ 
  $\beta_{21}$ & 2.00 & 1.94 & (1.74, 2.16) & 1.92 & (1.77, 2.09) \\ 
  $Z_1$ & 1.20 & 1.33 & (1.16, 1.53) & 1.23 & (1.08, 1.41) \\ 
  $Z_{21}$ & 0.50 & 0.51 & (0.28, 0.75) & 0.65 & (0.46, 0.87) \\ 
  $Z_{22}$ & 1.20 & 1.03 & (0.8, 1.37) & 0.96 & (0.69, 1.36) \\ 
  $Z_3$ & 1.00 & 0.86 & (0.7, 1.11) & 0.82 & (0.61, 1.15) \\ 
  $\phi_1$ & 5.00 & 5.44 & (3.85, 8.67) & 5.93 & (4.12, 8.48) \\ 
  $\phi_2$ & 25.00 & 17.36 & (10.7, 28.88) & 27.60 & (17.86, 45.99) \\ 
  $\tau^2$ & 0.50 & 0.47 & (0.27, 0.73) & 0.55 & (0.37, 0.8) \\  \hline
\end{tabular} 
\end{table}

\subsection{Application to LuftiBus-SNC Dataset}
\label{subsec:app}

In spatial epidemiology, the joint analysis of multiple diseases with similar etiology allows us to separate underlying risk factors into shared and disease-specific components. In this analysis, we examine the disease-specific spatial risk surface of lung cancer and shared spatial risk components between lung cancer and respiratory disease. 

Chronic lung disease contributes substantially to morbidity and mortality worldwide, with chronic obstructive pulmonary disease (COPD) being the third leading cause of death \citep{Lozano2012}. Forced expiratory volume in one second (FEV1) is a measure of the amount of air a person can exhale during a pulmonary test, it can be used to diagnose disease and predict respiratory-related mortality \citep{Menezes2014}. While respiratory disease and lung cancer share many common risk factors such as smoking and exposure to air pollution, it is of interest to examine the lung cancer specific spatial risk component. It may provide insights into identifying risk factors that are solely associated with lung cancer.

Initiated as a health promotion campaign by Lunge Zurich \citep{luftibus} in Switzerland, the `LuftiBus' project collected lung function measurements including FEV1 and demographic information from local residents. The data from LuftiBus observed between 2003 and 2012 were deterministically linked with a census-based Swiss National Cohort (SNC) study, to obtain 44,071 people with demographic, health and environmental variables in Switzerland. More importantly, the linkage provides us with the residential location of individual participants.

For lattice data, we compute the expected cause-specific (respiratory and lung cancer, respectively) mortalities in each municipality, adjusted by 5-year age-group and gender using the SNC data. We assume there are two latent spatial risk surfaces which are associated respiratory disease and lung cancer. The first risk surface is shared between FEV1, respiratory mortality and lung cancer mortality, while the second is lung cancer-specific risk surface. Typically with lattice data, multivariate conditional autoregressive models allow us to jointly analyze multiple responses and identify different latent components \citep{Jin2005}. However, municipal boundaries are artificial, we argue that a continuous spatial surface is a more natural modeling assumption. Therefore, we use our process-based unifying framework to conduct the analysis. Another advantage is that it allow us to incorporate the rich FEV1 data from Luftibus. The fusion model is structured as
\begin{align}
\boldsymbol{Y}_{\text{FEV1}} &\sim \text{N}(\beta_{1,0} + \beta_{1,1} \; X_\text{age} + \beta_{1,2} \; X_\text{gender} - Z_{11}\;w_1, \tau^2\boldsymbol{I}), \nonumber\\ \nonumber
\boldsymbol{Y}_{\text{resp}} &\sim \text{Pois}\bigl(E_{\text{resp}} \exp(\beta_{2,0} + {Z_{21}\; w_1})\bigr), \\ \nonumber
\boldsymbol{Y}_{\text{cancer}} &\sim \text{Pois}\bigl(E_{\text{cancer}} \exp(\beta_{3,0} + {Z_{31}\;w_1 + Z_{32}\; w_2}),\bigr) 
\end{align}
where $E_{\text{resp}}$ and $E_{\text{cancer}}$ are the expected cause-specific mortalities.

\begin{figure}
\includegraphics[width = \textwidth]{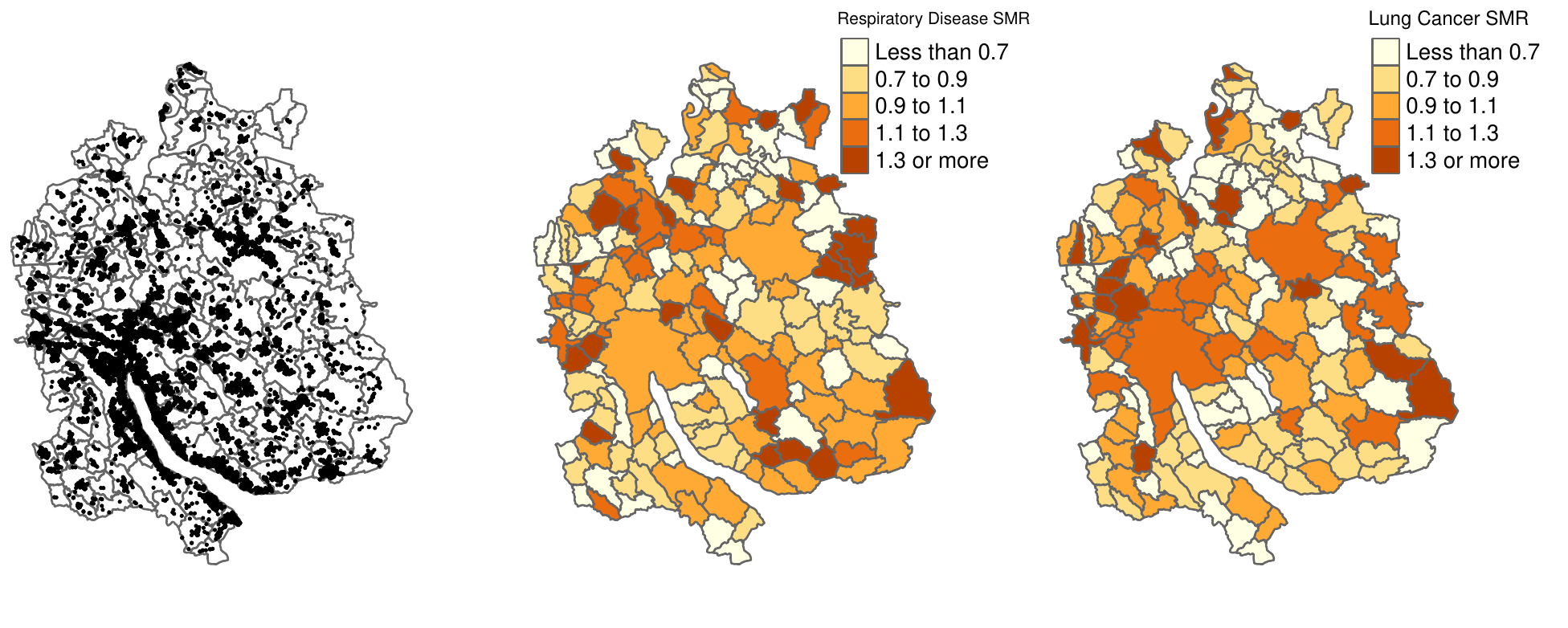}
\caption{Data used in the fusion model. Left: locations of geostatistical observation. Middle: respiratory standardized mortality ratio. Right: lung cancer standardized mortality ratio.}
\label{fig:dat}
\end{figure}

\begin{figure}
\includegraphics[width = \textwidth]{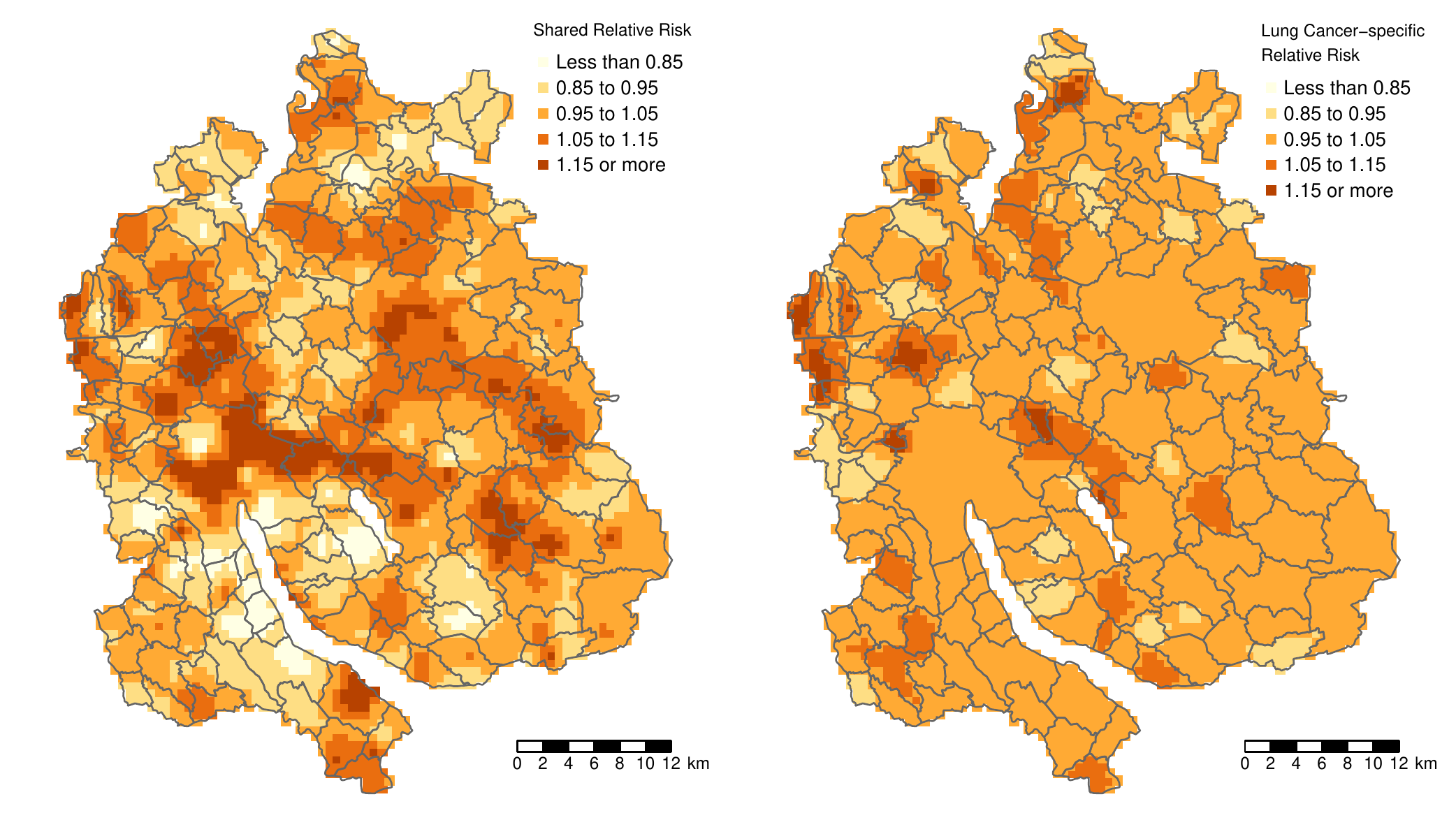}
\caption{Estimated spatial relative risk surfaces. Left: shared component between FEV1, respiratory mortality and lung cancer mortality. Right: lung cancer mortality-specific component.}
\label{fig:app}
\end{figure}

More than 60\% of the FEV1 measurements in the linked dataset are located in Canton of Zurich, therefore we restrict our analysis to Canton of Zurich. In addition, we focus the analysis on people who are 40 years or older, which results in 16,160 geostatistical observations. Since we have a large number of observations in the FEV1 outcome, the number of locations required to model in the latent processes is large. Therefore it is not feasible to use the Stan implementation. We conduct the analysis using the INLA implementation only. We use PC prior for the latent components with $\alpha = 1.5$ corresponding to exponential covariance function, median practical range of 1km and median $\sigma$ of 1. Figure~\ref{fig:dat} shows the locations of geostatistical observation and the standardized mortality ratio for respiratory disease and lung cancer. Figure~\ref{fig:app} shows the transformed posterior estimates of the latent processes representing relative risk surfaces in Canton of Zurich. The shared risk components between FEV1, respiratory mortality, and lung cancer mortality is highest in urban areas, with an effective range of 3.1 (95\% CI: 1.9, 5.2) km based on the exponential covariance function. The estimated relative risk is computed by exponentiating the latent process, which varies between 0.72 and 1.59. Meanwhile, the high-risk areas of lung cancer-specific components are scattered around Canton of Zurich, mainly in the north and west regions with an effective range of 1.2 (95\% CI: 0.3, 4.0) km. The variability is smaller than the shared component with values between 0.87 and 1.32, indicating a smoother risk surface compared to the shared component. The lung cancer-specific risk component is modeled via lattice data $\boldsymbol{Y}_\text{cancer}$ only, hence appearing to have some block-wise structures compared to the shared component.

\begin{table}[]
\caption{Parameter estimates and their 95\% posterior credible intervals (95\% CI) for the LuftiBus-SNC dataset. $\phi_1$ and $\phi_2$ are in meters.}
\label{tab:case}
\begin{tabular}{@{\extracolsep{8pt}}ccccc@{}}
\hline  \hline
Parameter & $\beta_{1,0}$ & $\beta_{1,1}$ & $\beta_{1,2}$ & $\beta_{2,0}$  \\ \hline
Median &  4.74 & 0.907 & -0.0375 & -0.0501 \\ 
95\% CI & (4.79, 4.7) & (0.923, 0.891) & (-0.0368, -0.0382) & (-0.101, -0.00127) \\   \hline\hline
Parameter & $\beta_{3,0}$ & $\tau^2$ & $\phi_1$ & $\phi_2$  \\ \hline
Median & -0.112 & 0.268 & 1020 & 389 \\  
95\% CI & (-0.171, -0.056) & (0.274, 0.262) & (628, 1720) & (87.3, 1320) \\ \hline\hline
Parameter & $Z_{11}$      & $Z_{21}$      &  $Z_{31}$     &  $Z_{32}$  \\ \hline
Median  & 0.0887 & 0.148 & 0.177 & 0.527 \\   
95\% CI   & (0.0223, 0.233) & (0.09, 0.257) & (0.0597, 0.449) & (0.242, 1.37)    \\ \hline
\end{tabular}
\end{table}

\section{Summary and Discussions}
\label{sec:dis}

We have proposed a unifying process-based statistical framework to handle spatial data fusion. The framework allows all three types of spatial data, namely geostatistical, lattice and point-pattern data, to be easily incorporated into a single multivariate spatial model. This framework contains theoretical and computational elements from several existing literatures: the basis for modeling latent processes in Stan is NNGP \citep{Datta2016}, the first Bayesian implementation is based on Stan \citep{rstan2016},  the alternative implementation uses INLA \citep{Rue2009}, the sampling point approximation approach for modeling lattice data is adopted from \cite{Fuentes2005}, discretization \citep{Moller1998} is used in modeling point-pattern data in Stan while data augmentation \cite{Simpson2016} is used in INLA. We have combined all of the individual elements and constructed this unifying framework. The framework extends upon existing flexible spatial fusion models \citep{Wang2018,Wilson2018} by making point-pattern data also compatible, hence completes all three spatial data types. We have benchmarked our INLA implementation against full Bayesian inference in Stan, and observed comparable performance with significantly decreased computation time. In addition, we have shown in the first simulation study that it is advantageous to conduct multivariate analysis using multiple spatial datasets if they are available.

Identifiability issues arise when there is more than one latent spatial process in the fusion model. Similar concern has been brought up in other multivariate spatial models \citep{Ren2013,Held2001}. Since the model becomes invariant under certain orthogonal transformations, the design matrix $\boldsymbol{Z}$ is not identifiable. \cite{Held2001} proposed a specific constraint on the relationship among the individual elements of the design matrix. \cite{Ren2013} proposed to constrain one element of each row in the design matrix to be strictly positive and having an ordered spatial range parameter. The same constraints allow identifiable parameters in the INLA implementation but not in Stan. A distinction between our proposed framework and existing multivariate models is, that we could potentially have at most one observation at any of the spatial locations even when we have three response variables. This makes identifying more than one latent process at each location problematic. Our INLA implementation does not directly model the latent variable parameters at the set of locations $\mathcal{U}$, but on the mesh vertices. One solution in the Stan implementation is to re-use the observed locations as sampling points and locations representing grid cells of LGCP whenever possible, such that the number of latent process parameters is reduced. Alternatively, certain elements of the design matrix can be constrained to zero based on expert knowledge, as done in our application to the LuftiBus-SNC dataset. When a model involves Mat\'{e}rn covariance function with smoothness parameter greater than 1, our implementation in Stan can be easily adapted by modifying the likelihood expressions, while INLA models can still be used as an approximation.

The usage of our proposed framework is multifaceted. The interest sometimes lies within latent spatial processes when analyzing spatial data, which represent residual spatial correlation in the response variables after taking existing covariates into consideration. The result can be used for detecting spatial clusters of unexplained risk or shared scientific drivers for response variables, which warrant further investigation in identifying those unknown drivers. When the interest is in predicting response variable for a newly observed spatial unit, fusion model improves the prediction of latent processes which in turn can improve the response variable prediction. Furthermore, the framework can be modified to use a one-dimensional Gaussian process in the latent components such that it applies beyond spatial data. For example, it can be used in time series modeling where all the observations are in $\mathcal{R}$ and some machine learning applications \citep{Rasmussen2005}.

Further research could be done on checking the compatibility of different data sources for spatial fusion modeling, i.e. if overlapping information exists between different spatial datasets. Such information can help to inform the model structure, especially the design matrix $\boldsymbol{Z}$. Needless to say, the framework can be extended to include temporal components.

\bigskip
\begin{center}
{\large\bf SUPPLEMENTARY MATERIAL}
\end{center}
All the R code used for the simulation studies is available in a separate file on the authors website. 

\bibliographystyle{chicago}
\bibliography{fusion_framework}

\begin{thebibliography}{}

\bibitem[\protect\citeauthoryear{Banerjee, Carlin, and Gelfand}{Banerjee
  et~al.}{2014}]{Banerjee2014}
Banerjee, S., B.~P. Carlin, and A.~E. Gelfand (2014).
\newblock {\em Hierarchical Modeling and Analysis for Spatial Data}.
\newblock CRC Press.
\newblock {p}. 136-139.

\bibitem[\protect\citeauthoryear{Banerjee, Gelfand, Finley, and Sang}{Banerjee
  et~al.}{2008}]{Banerjee2008}
Banerjee, S., A.~E. Gelfand, A.~O. Finley, and H.~Sang (2008).
\newblock {Gaussian predictive process models for large spatial data sets}.
\newblock {\em Journal of the Royal Statistical Society. Series B\/}~{\em
  70\/}(4), 825--848.

\bibitem[\protect\citeauthoryear{Berrocal, Gelfand, and Holland}{Berrocal
  et~al.}{2010}]{Berrocal2010}
Berrocal, V.~J., A.~E. Gelfand, and D.~M. Holland (2010).
\newblock A spatio-temporal downscaler for output from numerical models.
\newblock {\em Journal of Agricultural, Biological, and Environmental
  Statistics\/}~{\em 15\/}(2), 176--197.

\bibitem[\protect\citeauthoryear{Besag, York, and Molli{\'e}}{Besag
  et~al.}{1991}]{BYM}
Besag, J., J.~York, and A.~Molli{\'e} (1991).
\newblock {Bayesian Image Restoration, with Two Applications in Spatial
  Statistics}.
\newblock {\em The Annals of the Institute of Statistical Mathematics\/}~{\em
  43\/}(1), 1--20.

\bibitem[\protect\citeauthoryear{Brooks and Gelman}{Brooks and
  Gelman}{1998}]{Brooks1998}
Brooks, S.~P. and A.~Gelman (1998).
\newblock {General methods for monitoring convergence of iterative
  simulations}.
\newblock {\em Journal of Computational and Graphical Statistics\/}~{\em
  7\/}(4), 434--455.

\bibitem[\protect\citeauthoryear{Carpenter, Gelman, Hoffman, Lee, Goodrich,
  Betancourt, Brubaker, Guo, Li, and Riddell}{Carpenter
  et~al.}{2017}]{stan2017}
Carpenter, B., A.~Gelman, M.~Hoffman, D.~Lee, B.~Goodrich, M.~Betancourt,
  M.~Brubaker, J.~Guo, P.~Li, and A.~Riddell (2017).
\newblock Stan: {A} probabilistic programming language.
\newblock {\em Journal of Statistical Software\/}~{\em 76\/}(1).

\bibitem[\protect\citeauthoryear{Chammartin, Probst-Hensch, Utzinger, and
  Vounatsou}{Chammartin et~al.}{2016}]{Chammartin2016}
Chammartin, F., N.~Probst-Hensch, J.~Utzinger, and P.~Vounatsou (2016).
\newblock Mortality atlas of the main causes of death in {S}witzerland,
  2008-2012.
\newblock {\em Swiss Medical Weekly\/}~{\em 146}.

\bibitem[\protect\citeauthoryear{Cressie}{Cressie}{1991}]{cressie2015}
Cressie, N. (1991).
\newblock {\em Statistics for Spatial Data}.
\newblock John Wiley \& Sons.

\bibitem[\protect\citeauthoryear{Cressie and Johannesson}{Cressie and
  Johannesson}{2007}]{Cressie2007}
Cressie, N. and G.~Johannesson (2007).
\newblock Fixed rank kriging for very large spatial data sets.
\newblock {\em Journal of the Royal Statistical Society: Series B\/}~{\em
  70\/}(1), 209--226.

\bibitem[\protect\citeauthoryear{Datta, Banerjee, Finley, and Gelfand}{Datta
  et~al.}{2016}]{Datta2016}
Datta, A., S.~Banerjee, A.~O. Finley, and A.~E. Gelfand (2016).
\newblock Hierarchical nearest-neighbor {G}aussian process models for large
  geostatistical datasets.
\newblock {\em Journal of the American Statistical Association\/}~{\em
  111\/}(514), 800--812.

\bibitem[\protect\citeauthoryear{Follestad and Rue}{Follestad and
  Rue}{2003}]{follestad2003}
Follestad, T. and H.~Rue (2003).
\newblock {Modelling spatial variation in disease risk using Gaussian Markov
  random field proxies for Gaussian random fields}.
\newblock Technical report, Norwegian University of Science and Technology.

\bibitem[\protect\citeauthoryear{Fuentes and Raftery}{Fuentes and
  Raftery}{2005}]{Fuentes2005}
Fuentes, M. and A.~E. Raftery (2005).
\newblock Model evaluation and spatial interpolation by {B}ayesian combination
  of observations with outputs from numerical models.
\newblock {\em Biometrics\/}~{\em 61\/}(1), 36--45.

\bibitem[\protect\citeauthoryear{Fuglstad, Simpson, Lindgren, and Rue}{Fuglstad
  et~al.}{2018}]{Fuglstad2018}
Fuglstad, G.-A., D.~Simpson, F.~Lindgren, and H.~Rue (2018).
\newblock {Constructing Priors that Penalize the Complexity of Gaussian Random
  Fields}.
\newblock {\em Journal of the American Statistical Association\/}~{\em 0\/}(0),
  1--8.

\bibitem[\protect\citeauthoryear{Furrer, Genton, and Nychka}{Furrer
  et~al.}{2006}]{Furrer2006}
Furrer, R., M.~G. Genton, and D.~Nychka (2006).
\newblock Covariance tapering for interpolation of large spatial datasets.
\newblock {\em Journal of Computational and Graphical Statistics\/}~{\em
  15\/}(3), 502--523.

\bibitem[\protect\citeauthoryear{Gatrell, Bailey, Diggle, and
  Rowlingson}{Gatrell et~al.}{1996}]{Gatrell1996}
Gatrell, A.~C., T.~C. Bailey, P.~J. Diggle, and B.~S. Rowlingson (1996).
\newblock Spatial point pattern analysis and its application in geographical
  epidemiology.
\newblock {\em Transactions of the Institute of British Geographers\/}~{\em
  21\/}(1), 256--274.

\bibitem[\protect\citeauthoryear{Gelfand, Zhu, and Carlin}{Gelfand
  et~al.}{2001}]{Gelfand2001}
Gelfand, A.~E., L.~Zhu, and B.~P. Carlin (2001).
\newblock {On the change of support problem for spatio-temporal data}.
\newblock {\em Biostatistics\/}~{\em 2\/}(1), 31--45.

\bibitem[\protect\citeauthoryear{Gotway and Young}{Gotway and
  Young}{2002}]{Gotway2002}
Gotway, C.~A. and L.~J. Young (2002).
\newblock Combining incompatible spatial data.
\newblock {\em Journal of the American Statistical Association\/}~{\em
  97\/}(458), 632--648.

\bibitem[\protect\citeauthoryear{Greenland}{Greenland}{1992}]{Greenland1992}
Greenland, S. (1992).
\newblock Divergent biases in ecologic and individual-level studies.
\newblock {\em Statistics in Medicine\/}~{\em 11\/}(9), 1209--1223.

\bibitem[\protect\citeauthoryear{Homan and Gelman}{Homan and
  Gelman}{2014}]{Homan2014}
Homan, M.~D. and A.~Gelman (2014).
\newblock The {No-U-Turn Sampler}: Adaptively setting path lengths in
  {Hamiltonian Monte Carlo}.
\newblock {\em Journal of Machine Learning Research\/}~{\em 15\/}(1),
  1593--1623.

\bibitem[\protect\citeauthoryear{Jensen}{Jensen}{1906}]{jensen1906}
Jensen, J. L. W.~V. (1906).
\newblock Sur les fonctions convexes et les inégalités entre les valeurs
  moyennes.
\newblock {\em Acta Mathematica\/}~{\em 30}, 175--193.

\bibitem[\protect\citeauthoryear{Jin, Carlin, and Banerjee}{Jin
  et~al.}{2005}]{Jin2005}
Jin, X., B.~P. Carlin, and S.~Banerjee (2005).
\newblock Generalized hierarchical multivariate {CAR} models for areal data.
\newblock {\em Biometrics\/}~{\em 61\/}(4), 950--961.

\bibitem[\protect\citeauthoryear{Kelsall and Wakefield}{Kelsall and
  Wakefield}{2002}]{Kelsall2002}
Kelsall, J. and J.~Wakefield (2002).
\newblock Modeling spatial variation in disease risk: A geostatistical
  approach.
\newblock {\em Journal of the American Statistical Association\/}~{\em
  97\/}(459), 692--701.

\bibitem[\protect\citeauthoryear{Knorr-Held and Best}{Knorr-Held and
  Best}{2001}]{Held2001}
Knorr-Held, L. and N.~G. Best (2001).
\newblock A shared component model for detecting joint and selective clustering
  of two diseases.
\newblock {\em Journal of the Royal Statistical Society. Series A\/}~{\em
  164\/}(1), 73--85.

\bibitem[\protect\citeauthoryear{Krainski, {G{\'o}mez Rubio}, Bakka, Lenzi,
  Castro-Camilo, Simpson, Lindgren, and Rue}{Krainski et~al.}{2018}]{inlabook}
Krainski, E., V.~{G{\'o}mez Rubio}, H.~Bakka, A.~Lenzi, D.~Castro-Camilo,
  D.~Simpson, F.~Lindgren, and H.~Rue (2018).
\newblock {\em {Advanced Spatial Modeling with Stochastic Partial Differential
  Equations Using R and INLA}}.
\newblock Chapman and Hall/CRC.

\bibitem[\protect\citeauthoryear{Kyriakidis, Kim, and Miller}{Kyriakidis
  et~al.}{2001}]{Kyriakidis2001}
Kyriakidis, P.~C., J.~Kim, and N.~L. Miller (2001).
\newblock Geostatistical mapping of precipitation from rain gauge data using
  atmospheric and terrain characteristics.
\newblock {\em Journal of Applied Meteorology\/}~{\em 40\/}(11), 1855--1877.

\bibitem[\protect\citeauthoryear{Lindgren and Rue}{Lindgren and
  Rue}{2015}]{rINLA}
Lindgren, F. and H.~Rue (2015).
\newblock {Bayesian spatial modelling with R-INLA}.
\newblock {\em Journal of Statistical Software, Articles\/}~{\em 63\/}(19),
  1--25.

\bibitem[\protect\citeauthoryear{Lindgren, Rue, and Lindström}{Lindgren
  et~al.}{2011}]{Lindgren2011}
Lindgren, F., H.~Rue, and J.~Lindström (2011).
\newblock {An explicit link between Gaussian fields and Gaussian Markov random
  fields: the stochastic partial differential equation approach}.
\newblock {\em Journal of the Royal Statistical Society: Series B\/}~{\em
  73\/}(4), 423--498.

\bibitem[\protect\citeauthoryear{Liu, Le, and Zidek}{Liu
  et~al.}{2011}]{Liu2011}
Liu, Z., N.~D. Le, and J.~V. Zidek (2011).
\newblock An empirical assessment of {B}ayesian melding for mapping ozone
  pollution.
\newblock {\em Environmetrics\/}~{\em 22\/}(3), 340--353.

\bibitem[\protect\citeauthoryear{Lozano and et~al}{Lozano and
  et~al}{2012}]{Lozano2012}
Lozano, R. and et~al (2012).
\newblock {Global and regional mortality from 235 causes of death for 20 age
  groups in 1990 and 2010: a systematic analysis for the Global Burden of
  Disease Study 2010}.
\newblock {\em The Lancet\/}~{\em 380\/}(9859), 2095--2128.

\bibitem[\protect\citeauthoryear{{Lunge Z\"{u}rich}}{{Lunge
  Z\"{u}rich}}{2017}]{luftibus}
{Lunge Z\"{u}rich} (2017).
\newblock The {LuftiBus} {P}roject.
\newblock {A}ccessed: 2017--02--28.

\bibitem[\protect\citeauthoryear{MacNab}{MacNab}{2016}]{MacNab2016}
MacNab, Y.~C. (2016).
\newblock Linear models of coregionalization for multivariate lattice data: a
  general framework for coregionalized multivariate {CAR} models.
\newblock {\em Statistics in Medicine\/}~{\em 35\/}(21), 3827--3850.

\bibitem[\protect\citeauthoryear{Martins, Simpson, Lindgren, and Rue}{Martins
  et~al.}{2013}]{Martins2013}
Martins, T.~G., D.~Simpson, F.~Lindgren, and H.~Rue (2013).
\newblock {Bayesian computing with INLA: New features}.
\newblock {\em Computational Statistics \& Data Analysis\/}~{\em 67}, 68--83.

\bibitem[\protect\citeauthoryear{McMillan, Holland, Morara, and Feng}{McMillan
  et~al.}{2010}]{McMillan2010}
McMillan, N.~J., D.~M. Holland, M.~Morara, and J.~Feng (2010).
\newblock Combining numerical model output and particulate data using
  {B}ayesian space–time modeling.
\newblock {\em Environmetrics\/}~{\em 21\/}(1), 48--65.

\bibitem[\protect\citeauthoryear{Menezes, P{\'e}rez-Padilla, Wehrmeister,
  Lopez-Varela, Mui{\~n}o, Valdivia, Lisboa, Jardim, de~Oca~Maria, Talamo,
  Bielemann, Gazzotti, Laurenti, Celli, Victora, and for~the
  PLATINO~team}{Menezes et~al.}{2014}]{Menezes2014}
Menezes, A. M.~B., R.~P{\'e}rez-Padilla, F.~C. Wehrmeister, M.~V. Lopez-Varela,
  A.~Mui{\~n}o, G.~Valdivia, C.~Lisboa, J.~R.~B. Jardim, M.~de~Oca~Maria,
  C.~Talamo, R.~Bielemann, M.~Gazzotti, R.~Laurenti, B.~Celli, C.~G. Victora,
  and for~the PLATINO~team (2014).
\newblock {Lozano2012}.
\newblock {\em PLOS ONE\/}~{\em 9\/}(10), 1--10.

\bibitem[\protect\citeauthoryear{M{\o}ller, Syversveen, and
  Waagepetersen}{M{\o}ller et~al.}{1998}]{Moller1998}
M{\o}ller, J., A.~R. Syversveen, and R.~P. Waagepetersen (1998).
\newblock {Log Gaussian Cox processes}.
\newblock {\em Scandinavian Journal of Statistics\/}~{\em 25\/}(3), 451--482.

\bibitem[\protect\citeauthoryear{Moraga, Cramb, Mengersen, and Pagano}{Moraga
  et~al.}{2017}]{Moraga2017}
Moraga, P., S.~M. Cramb, K.~L. Mengersen, and M.~Pagano (2017).
\newblock {A geostatistical model for combined analysis of point-level and
  area-level data using INLA and SPDE}.
\newblock {\em Spatial Statistics\/}~{\em 21}, 27--41.

\bibitem[\protect\citeauthoryear{Ogata}{Ogata}{1988}]{Ogata1998}
Ogata, Y. (1988).
\newblock Statistical models for earthquake occurrences and residual analysis
  for point processes.
\newblock {\em Journal of the American Statistical Association\/}~{\em
  83\/}(401), 9--27.

\bibitem[\protect\citeauthoryear{{R Core Team}}{{R Core Team}}{2018}]{R2017}
{R Core Team} (2018).
\newblock {\em R: A Language and Environment for Statistical Computing}.
\newblock Vienna, Austria: R Foundation for Statistical Computing.

\bibitem[\protect\citeauthoryear{Rasmussen and Williams}{Rasmussen and
  Williams}{2005}]{Rasmussen2005}
Rasmussen, C.~E. and C.~K.~I. Williams (2005).
\newblock {\em Gaussian Processes for Machine Learning (Adaptive Computation
  and Machine Learning)}.
\newblock The MIT Press.

\bibitem[\protect\citeauthoryear{Ren and Banerjee}{Ren and
  Banerjee}{2013}]{Ren2013}
Ren, Q. and S.~Banerjee (2013).
\newblock {Hierarchical factor models for large spatially misaligned data: a
  low-rank predictive process approach}.
\newblock {\em Biometrics\/}~{\em 69\/}(1), 19--30.

\bibitem[\protect\citeauthoryear{Rue, Martino, and Chopin}{Rue
  et~al.}{2009}]{Rue2009}
Rue, H., S.~Martino, and N.~Chopin (2009).
\newblock Approximate {B}ayesian inference for latent {G}aussian models by
  using integrated nested {L}aplace approximations.
\newblock {\em Journal of the Royal Statistical Society: Series B\/}~{\em
  71\/}(2), 319--392.

\bibitem[\protect\citeauthoryear{Sahu, Gelfand, and Holland}{Sahu
  et~al.}{2010}]{Sahu2010}
Sahu, S.~K., A.~E. Gelfand, and D.~M. Holland (2010).
\newblock Fusing point and areal level space-time data with application to wet
  deposition.
\newblock {\em Journal of the Royal Statistical Society Series C\/}~{\em
  59\/}(1), 77--103.

\bibitem[\protect\citeauthoryear{Schmidt and Gelfand}{Schmidt and
  Gelfand}{2003}]{Schmidt2003}
Schmidt, A.~M. and A.~E. Gelfand (2003).
\newblock {A Bayesian coregionalization approach for multivariate pollutant
  data}.
\newblock {\em Journal of Geophysical Research: Atmospheres\/}~{\em
  108\/}(D24).

\bibitem[\protect\citeauthoryear{Shi and Kang}{Shi and Kang}{2017}]{Shi2017}
Shi, H. and E.~L. Kang (2017).
\newblock Spatial data fusion for large non-{Gaussian} remote sensing datasets.
\newblock {\em Stat\/}~{\em 6\/}(1), 390--404.

\bibitem[\protect\citeauthoryear{Simpson, Illian, Lindgren, Sørbye, and
  Rue}{Simpson et~al.}{2016}]{Simpson2016}
Simpson, D., J.~B. Illian, F.~Lindgren, S.~H. Sørbye, and H.~Rue (2016).
\newblock {Going off grid: computationally efficient inference for log-Gaussian
  Cox processes}.
\newblock {\em Biometrika\/}~{\em 103\/}(1), 49--70.

\bibitem[\protect\citeauthoryear{{Stan Development Team}}{{Stan Development
  Team}}{2018}]{rstan2016}
{Stan Development Team} (2018).
\newblock {RStan}: the {R} interface to {Stan}.
\newblock {R} package version 2.18.0.

\bibitem[\protect\citeauthoryear{Stein}{Stein}{2008}]{Stein2008}
Stein, M.~L. (2008).
\newblock {A modeling approach for large spatial datasets}.
\newblock {\em Journal of the Korean Statistical Society\/}~{\em 37\/}(1),
  3--10.

\bibitem[\protect\citeauthoryear{Wackernagel}{Wackernagel}{2003}]{Wackernagel2003}
Wackernagel, H. (2003).
\newblock {\em {Multivariate Geostatistics: An Introduction with
  Applications}}.
\newblock Berlin, Heidelberg: Springer Berlin Heidelberg.

\bibitem[\protect\citeauthoryear{Wang and Furrer}{Wang and
  Furrer}{2019}]{Wang2019}
Wang, C. and R.~Furrer (2019).
\newblock Efficient inference of generalized spatial fusion models with
  flexible specification.
\newblock {\em Stat\/}~{\em 8\/}(1), e216.

\bibitem[\protect\citeauthoryear{Wang, Puhan, and Furrer}{Wang
  et~al.}{2018}]{Wang2018}
Wang, C., M.~A. Puhan, and R.~Furrer (2018).
\newblock {Generalized spatial fusion model framework for joint analysis of
  point and areal data}.
\newblock {\em Spatial Statistics\/}~{\em 23}, 72--90.

\bibitem[\protect\citeauthoryear{Wang and Wall}{Wang and Wall}{2003}]{Wang2003}
Wang, F. and M.~M. Wall (2003).
\newblock {Generalized common spatial factor model}.
\newblock {\em Biostatistics\/}~{\em 4\/}(4), 569--582.

\bibitem[\protect\citeauthoryear{Wilson and Wakefield}{Wilson and
  Wakefield}{2018}]{Wilson2018}
Wilson, K. and J.~Wakefield (2018).
\newblock {Pointless spatial modeling}.
\newblock {\em Biostatistics\/}, 1--16.

\bibitem[\protect\citeauthoryear{Zhang, Datta, and Banerjee}{Zhang
  et~al.}{2019}]{zhang2019}
Zhang, L., A.~Datta, and S.~Banerjee (2019).
\newblock Practical bayesian modeling and inference for massive spatial data
  sets on modest computing environments†.
\newblock {\em Statistical Analysis and Data Mining: The ASA Data Science
  Journal\/}~{\em 12\/}(3), 197--209.

\end{thebibliography}

\end{document}